# Coupling model analysis of interchain coupled chain dynamics of PEO in blends with PMMA

K. L. Ngai[1,2] and Li-Min Wang[2]

[1]*CNR-IPCF, Dipartimento di Fisica, Universita di Pisa, Pisa, Italy*
[2]*State Key Lab of Metastable Materials Science and Technology, Yanshan University, Qinhuangdao, Hebei, China*

**Abstract**

Quasielastic neutron scattering and molecular dynamics simulation data from PEO/PMMA blends found that for short times the self-dynamics of PEO chain follows the Rouse model, but at longer times past $t_c=1$ to 2 ns it becomes slower and departs from the Rouse model in dependences on time, momentum transfer, and temperature. To explain the anomalies, others had proposed the random Rouse model (RRM) in which each monomer has different mobility taken from a broad log-normal distribution. Despite the success of the RRM, Diddens, Brodeck and Heuer [EPL, **95**, 56003 (2011)] extracted the distribution of friction coefficients from the MD simulations of a PEO/PMMA blend and found the distribution is much narrower than expected from the RRM. We propose a simpler alternative explanation of the data by utilizing alone the observed crossover of PEO chain dynamics at $t_c$. The present problem is just a special case of a general property of relaxation in interacting systems, which is the crossover from independent relaxation to coupled many-body relaxation at some $t_c$ determined by the interaction potential. The generality is brought out vividly by pointing out that the crossover also had been observed by neutron scattering from entangled chains relaxation in monodisperse homo-polymers, and from the segmental α-relaxation of PEO in blends with PMMA. The properties of all the relaxation processes in connection with the crossover are similar, despite the length-scales of the relaxation in these systems are widely different.

## 1. Introduction

The dynamics of poly(ethylene oxide) (PEO) in blends rich in poly(methyl methacrylate) (PMMA) have been studied at various length scales including local secondary β-relaxation, segmental α-relaxation responsible for glass transition of the PEO component [1-4], and unentangled PEO chain relaxation and diffusion [5-8]. The interest of the research community on the so-called asymmetric blends, such as the $x$PEO/(1-$x$)PMMA, is because of the exceedingly large difference in the glass transition temperature $T_g$ of the two components. More recent study of this asymmetric blend by using quasielastic neutron scattering has found anomalous global dynamics of the unentangled PEO chains [6]. At short times less than about 1 ns, the dynamics of PEO chain follow the Rouse model. However, at longer times the PEO chains dynamics becomes much slower. Based on the average Rouse relaxation rates obtained from the data less than 1 ns, the Rouse model fails at time longer than 1 ns by predicting a much faster decay than observed. The data show no evidence for a characteristic length scale to indicate confinement of the PEO chain by the much less mobile PMMA chains. To explain the anomalous PEO chain dynamics in the blend, the random Rouse model (RRM) was introduced [6]. In this model, each repeat unit has its own mobility obeying a broad Gaussian distribution, and global chain dynamics of the PEO component is not determined by the average local environment alone. A fully atomistic molecular dynamics (MD) simulation [7] has confirmed the results of the quasielastic neutron scattering study, and provide more information and results on the PEO chain self-motion.

A newly developed simulation method was used to extract the distribution of friction coefficients from molecular dynamics simulations of a PEO/PMMA blend to test the RRM explicitly [8]. The distribution was found to be much narrower than the RRM needs to explain the data, and thus it casts doubt on the RRM explanation of the anomalous PEO chain dynamics. Instead it was suggested that the presence of additional forward-backward correlations of intermolecular origin is responsible for the anomalous PEO behavior [8]. This finding and the suggestion of intermolecular involvement in the PEO chain relaxation encourage the search for alternative explanations by models based on intermolecular coupling. One candidate is the Coupling Model CM) [9-16]. In this paper, the CM is used to address the experimental data from neutron scattering study and the MD simulations. More characteristics of the unusual PEO chain dynamics in the blend obtained by MD simulations provide more ground for applying the predictions of the CM for explanation.

Before proceeding any further, it is important to qualify the statement made above that could be misconstrued as if the Rouse model is totally valid in homo-polymer or in polymer blends. Actually, the dynamics of PEO chain follow the Rouse model predictions only in the properties probed by neutron scattering in unentangled homo-polymers such as PEO [17], and PEO in blends with PMMA [6], and the same properties were confirmed by MD simulations [7,17]. There are other finer properties of the Rouse model that are at variance with the chain dynamics of unentangled homo-polymers found by MD simulations and experiments [17-23], and theory [23-25]. These deviations are caused by the presence of interchain coupling in bulk polymers, which is not included in the Rouse model. Therefore, from now on, whenever the Rouse model is mentioned, it is restricted to the properties that are in accord with quasielastic neutron scattering [6,26,27] and MD simulations [7,17] studies of $x$PEO/(1-$x$)PMMA in the temperature and momentum transfer ranges and at time shorter than 1 to 2 ps.

## 2. Crossover from Rouse dynamics to intermolecular coupled chain dynamics at $t_c \approx 1$ ns

The Rouse model has the mean-square displacement, $<r^2(t)>$, of a chain segment increases proportionally to the square root of time according to

$$<r^2(t)> = 2(W(T)l^4 t/\pi)^{0.5} \qquad (1)$$

where $W(T) = 3k_B T/\zeta(T)l^2$ is the elementary Rouse relaxation rate, $\zeta(T)$ is the friction coefficient, and $l$ the segment length [6,7]. In the Gaussian approximation the segmental self-correlation function relates directly to $<r^2(t)>$, resulting in the intermediate self-dynamic structure factor having the Gaussian form

$$S(Q,t) = \exp\left[-Q^2 <r^2(t)>/6\right] = \exp[-(Q^2 W^{0.5} l^4 \pi^{-0.5}/3) t^{0.5}] \qquad (2)$$

The time dependence is the stretched exponential function of time, $\exp[-(t/\tau_R)^\beta]$, with $\beta$=0.5. The experiments performed at the backscattering instruments by Niedzwiedz et al. [6] are capable of measuring the self-motion of PEO chains up to 1 ns. The mean square displacements of PEO chains in 35%PEO/65%PMMA sample were obtained by Fourier transformation of backscattering spectra at $Q$=2.4 and 3.2 nm$^{-1}$ and temperatures from 350 to 400 K. The results show the PEO chain dynamics are in agreement with the Rouse model prediction given by Eq.(2) for $t \leq 1$ ns.

Additional measurements of the PEO chain motion in the PMMA matrix up to 80 ns were made using the neutron spin echo spectrometers. The measured PEO single chain dynamic structure factor, $S(Q,t)/S(Q)$, for all small momentum transfer $Q$ from 1 to 3 nm$^{-1}$ shows its decay is much slower than that expected from the Rouse rate determined at times shorter than 1ns by the backscattering instruments. The slowing down was quantified by fitting to an effective smaller Rouse rate. The fits show retardations by factors of 4 to 20, depending on PEO content and temperature. Moreover, the time dependence of $S(Q,t)/S(Q)$ is at variance with the Rouse model. The crossover time from Rouse behavior to retarded and non-Rousean behavior determined by the neutron scattering experiments is about 1 ns, independent of temperature $T$ and $Q$.

MD simulations have obtained the incoherent scattering functions $F_s(Q,t)$ of PEO chains in the blend for $0.1<Q<0.4$ Å$^{-1}$, $300\leq T\leq 500$ K up to almost 100 ns. In the same $Q$, $T$, and time ranges, Rouse behavior was found before for pure PEO [17] as evidenced by the stretched exponential time dependence, $\exp[-(t/\tau)^\beta]$, of $F_s(Q,t)$ with $\beta = 0.5$. In order to compare with neutron scattering experimental results, $F_s(Q,t)$ of PEO chains in the blend was considered in the restricted time range from 60 ps to 2 ns corresponding roughly to the dynamic range of the backscattering instruments. It was found in this restricted time range the stretched exponential function with $\beta = 0.5$ is a good description of $F_s(Q,t)$ (see Fig.2 in Ref.[7]), and the averaged $\tau$ has the $Q^{-4}$-dependence (see Fig.4b in Ref.7). All these features are characteristics of Rouse dynamics, and thus verifying that Rouse dynamics holds for $t<2$ ns, again independent of $T$ and $Q$. Notwithstanding, the $\beta=0.5$ fit of the restricted time range does not describe the full time decay of $F_s(Q,t)$, which is more stretched and retarded in the longer time range, consistent with the finding of neutron scattering results.

The neutron scattering and MD simulation data of the unentangled PEO chains in 35%PEO/65%PMMA summarized in the above clearly demonstrate at times shorter than about 1 ns that the relaxation is describable by the Rouse model, independent of $T$ and $Q$ chosen for the studies. However, beyond 1 ns, it is slowed down and departs from Rouse dynamics. Hence there is crossover in the dynamics of PEO chains at the crossover time $t_c\approx 1$ to 2 ns. Similar crossover of chain dynamics had been found in entangled poly(dimethylsiloxane) (PDMS) [26] and alternating copolymer of poly(ethylene propylene) (PEP) at comparable value of the crossover time $t_c\approx 1$ ns, and it is also independent of $T$ and $Q$. [27]. The physics governing unentangled PEO in blend with PMMA is obviously different from the entangled PDMS and PEP because the latter are entangled and nearly monodisperse polymers, with all chains equivalent. The fundamental aspect shared by the two systems is interchain interaction, which makes the crossover of dynamics so similar. The core of the Coupling Model for interacting many-body systems is the crossover from primitive or independent relaxation to slowed-down coupled relaxation at some crossover time $t_c$. The crossover is caused by the onset of classical chaos in the phase space originating from the nonlinear (anharmonic) interaction potential. The onset time is determined by the strength of the interaction [9-12,16], and therefore the crossover time $t_c$ are independent of $Q$ and $T$, as found by experiments and simulations. This is a general feature found experimentally in different kinds of interacting systems and relaxation processes even outside the realm of polymers [16]. The reason why $t_c\approx 1$ ns for unentangled PEO chains in blends as well as for entangled PDMS and PEP is because inter-chain interaction is common to all of these systems. However, for the segmental α-relaxation in polymers, the crossover from primitive or independent relaxation to intermolecularly coupled α-relaxation has been found to occur at a much shorter time of $t_c\approx 2$ ps [28-32]. Interaction between monomers executing the segmental α-relaxation involves length-scales much shorter than between chains, and hence interaction is much stronger in the former than the latter. In general, $t_c$ becomes longer on weakening the interaction. In the limit of vanishing interaction, the primitive relaxation prevails at all times and $t_c\to\infty$. Thus, the fact that $t_c\approx 1$ ns observed for global chain dynamics is much longer than $t_c\approx 2$ ps for segmental α-relaxation can be rationalized. Inter-chain interaction is further reduced on adding a solvent to polymer to separate the chains apart but they are still entangled in semi-dilute solutions, and in fact crossover from Rouse dynamics to non-Rousean dynamics at $t_c\approx 1$ μs or longer was observed in semidilute polymers solutions and related systems [33-37].

The original Rouse model in dilute solution generalized to undiluted polymers and blends [38] basically still involves the motion of a single chain without taking into account of inter-chain interaction. It is the primitive chain motion in the context of the CM present at short times $t\leq t_c\approx 1$ ns. After $t_c\approx 1$ ns, inter-chain interaction present in undiluted polymers and blends slows down the Rouse dynamics as well as modifying their time dependence. It is worthwhile to reemphasize that the observation of the anomalous PEO chain dynamics in blends with PMMA by neutron scattering [6] and by simulations [7] is just another manifestation of the crossover from primitive relaxation to many-body coupled relaxation, a general property of relaxation in interacting systems. In the following section, we demonstrate the crossover and the consequences associated with it are sufficient to explain the data from neutron scattering experiment and MD simulations. There is no need to introduce any additional element such as the log-normal distribution of monomeric friction coefficients in the RRM to explain the data. This is one of the major conclusions of this paper.

More characteristics of the PEO chain dynamics were found by MD simulations than by neutron scattering. In the following section we address these properties of the slowed-down and non-Rousean dynamics for $t>t_c$, and their relations to the Rouse dynamics for $t<t_c$, of PEO in blends. Rationalization of these properties and relations is provided by the CM.

## 3. Anomalous properties of the PEO chain in blends

MD simulations of the normalized self-correlation functions $F_s(Q,t)$ have found that two different dynamics of PEO chains in blends at times separated by $t_c \approx 2$ ns. The slowed-down and non-Rousean dynamics appear only after 2 ps and considered in Ref.[7] up o 20 ns. This limited time range from 2 ns to 20 ns alone makes it difficult to characterize its dynamics. This is the reason why Brodeck et al. [7] choose to fit $F_s(Q,t)$ in the broader time range from 2 ps to 20 ns to the stretched exponential functions,

$$F_s(Q,T) = \exp\{-[t/\tau(Q,T)]^{-\beta(Q,T)}\} \tag{3}$$

with $\beta(T,Q)$ and $\tau(T,Q)$ as free parameters for all $Q$ and $T$. Since two different regimes of dynamics are mixed together in the fit, the parameters of the fit, $\beta(T,Q)$ and $\tau(T,Q)$, do not exactly characterize the slowed-down and non-Rousean dynamics after 2 ns. The observed trend of the decrease of $\beta(T,Q)$ on decreasing $T$ is reflection of the increasing contribution to $F_s(Q,t)$ from the non-Rousean dynamics. Naturally the fit underestimates the rate of decay of $F_s(Q,t)$ at times shorter than 1 ns especially at higher temperatures, because the Rouse model applies there, and the stretch exponential has $\beta=0.5$. Notwithstanding, the fits adequately account the time dependence of $F_s(Q,t)$ from 2 to 20 ns, as can be verified by inspection of Fig.3a of Ref.[7], and hence $\beta(T,Q)$ and $\tau(T,Q)$ can be accepted as acceptable approximations for the non-Rousean dynamics.

Longer $\tau(T,Q)$ fares better as approximation because the overall fit puts more emphasis on the slower relaxation. Therefore, we consider first the data of $\beta(T,Q)$ and $\tau(T,Q)$ given in Figs.4a, 4b, 4c and 5a of Ref.[7] at the lower $Q$ values of 0.1, 0.2, and 0.3 Å$^{-1}$, and the lower temperatures of 300 and 350 K, the values of which are all significantly longer than 2 ns. The $Q^{-\gamma}$-dependence of $\tau(T,Q)$ at 300 and 350 K in Fig.4b with $\gamma=6.6$ and 5.6 respectively Fig.4b deviates greatly from $\gamma=4$ of Rouse dynamics which is valid for $t\leq t_c\approx 1$ ns. Brodeck et al. have shown the anomalous $Q^{-\gamma}$-dependence of $\tau(T,Q)$ at 300 and 350 K can be explained by assuming Gaussian behavior of PEO chain. To do this, Eq.(3) is now rewritten to display explicitly the $Q^2$ term from Gaussian behavior. From this form,

$$F_s(Q,T) = \exp\{-Q^2[(t/\tau(Q,T)]^{-\beta(Q,T)}\}. \tag{4}$$

By inserting the $Q^2$ term back inside the square brackets, it follows that

$$\tau(Q,T) \propto Q^{-2/\beta(T,Q)}. \tag{5}$$

This easy derivation of the $Q^{-\gamma}$-dependence of $\tau(T,Q)$ from Gaussian behavior alone results in having $\gamma=2/\beta(T,Q)$, in agreement with MD simulations. The values of $\beta(T,Q)$ at 300 and 350 K are 0.30 and 0.34 respectively, and the corresponding values of $\gamma=2/\beta(T,Q)$ are 0.67 and 0.59 is good agreement with the observed $\gamma=6.6$ and 5.6. Notwithstanding, from the non-Gaussianity parameter reported [7] it is not clear that the assumption of Gaussian behavior is valid at times longer than 2 ps. Moreover and despite the ease in explaining the anomalous $Q^{-\gamma}$-dependence of $\tau(T,Q)$ by the assumption of Gaussian behavior in Ref.[7], no the temperature dependence had not been addressed. Observed by MD simulations is the much stronger temperature dependence of $\tau(T,Q)$ for the slowed-down and non-Rousean dynamics at times after 2 ns than that of the Rouse relaxation time $\tau_R(T,Q)$ obtained by the fit to Eq.(2) restricted to times shorter than 2 ns. We made the comparison quantitative by extracting the increase of $\tau(T,Q)$ and of $\tau_R(T,Q)$ on decreasing temperature from 500 down to 300 K at $Q = 0.2$, and 0.3 Å$^{-1}$. The increase of $\tau(T,Q)$ and of $\tau_R(T,Q)$ at $Q=0.3$ Å$^{-1}$ is about 3.7 decades and 2.2 decades respectively, and at $Q=0.2$ Å$^{-1}$ the increase is about 4.1 decades and 2.2 decades respectively. The same was carried out on decreasing temperature from 500 K to 350 K. There, the increase of $\tau(T,Q)$ and of $\tau_R(T,Q)$ at $Q=0.3$ Å$^{-1}$ is about 2.2 decades and 1.3 decades respectively, at $Q=0.2$ Å$^{-1}$ the increase is about 2.4 decades and 1.3 respectively, and at $Q=0.1$ Å$^{-1}$ the increase is about 3.0 decades and 1.3 decades respectively.

To explain simultaneously the stronger $Q^{-\gamma}$-dependence and the enhanced temperature dependence of $\tau(T,Q)$, we use the crossover of the self-correlation functions $F_s(Q,t)$ at $t_c \approx 2$ ns from Rouse dynamics observed for for $t \leq 1$ ns and given by Eq.(2) to the slowed-down and non-Rouse dynamics represented by Eq.(3). The crossover is not sharp at $t_c$ but likely over some small neighborhood of time where the function and its derivative undergo continuous change. Nevertheless, without exact information on the time dependence of $F_s(Q,t)$ on crossing $t_c \approx 2$ ns, we make the approximation of equating the two expressions for $F_s(Q,t)$ at $t_c$. From this, we simultaneously obtained the dependence on $Q$ and $T$ of $\tau(T,Q)$ given by

$$\tau(T,Q) = CQ^{-[\frac{2}{\beta(T,Q)}]}W(T)^{-[\frac{0.5}{\beta(T,Q)}]}, \qquad (6)$$

where $C$ is a proportionality constant. Since $\beta(T,Q)$ is smaller than 0.5 for all $T$ and $Q$ at which $\tau(T,Q)$ is significantly longer than 2 ns, the exponent of $W(T)$ is larger than one. Hence the temperature dependence of $\tau(T,Q)$ is stronger than $\tau_R(T,Q)$ as shown in Fig.4b of Ref.[7], and in the preceding paragraph. Inter-chain coupling of PEO chain in the blend is expected to be enhanced with increasing concentration of the less mobile PMMA. In the context of the CM, this leads to an increase of the coupling parameter, $n(T,Q) \equiv [1-\beta(T,Q)]$, or a decrease of $\beta(T,Q)$, and hence a broader dispersion and a stronger temperature dependence of $\tau_R(T,Q)$ from Eq.(6). These expected changes on increasing PMMA content in the blend have been found by neutron scattering experiments as hinted in Ref.[6] by the need to increase the width of distribution of friction coefficients to fit the data by the random Rouse Model.

By Eq.(6), we can explain not only the anomalous $Q^{-\gamma}$-dependence of $\tau(T,Q)$ with $\gamma$ roughly equal to $2/\beta(T,Q)$, but also the stronger temperature dependence of $\tau(T,Q)$ (see Figs.4b and 5a) found by MD simulations. Explanation of the latter is nontrivial particularly without added assumption, as done here by a natural and direct consequence of the crossover from primitive Rouse dynamics of the PEO chains in the blend to inter-chain coupled PEO chain relaxation. Conventional models and theory of polymer dynamics have only a single friction coefficient governing all relaxation and diffusion processes, and thus cannot explain the data without added assumptions such as the broad distribution of friction coefficients in the RRM.

The experimentally observed crossover in the presence case is just one special case of the general phenomenon observed in other relaxation and diffusion processes in different interacting systems [26-37]. The crossover is also the major feature of the CM that has been derived for some simplified models [9-12]. The benefit that the crossover brings to the presence case is simultaneous accounting of the anomalous $Q$-dependence and the enhanced temperature dependence of the slowed-down coupled PEO chain relaxation. The benefit of multiple predictions in successfully explaining experimental data has been realized in other relaxation processes and different systems [16]. Restricting to problems related to polymer chain dynamics, we can cite several examples where the crossover had been found directly by experiment [26,27]. Applying the consequence of the crossover to the entangled linear and star branched polyethylene and hydrogenated polybutadiene, we had explained simultaneously the molecular weight dependence and the Arrhenius activation energy of the viscosity and diffusion coefficient [13,15,16]. For semidilute polymer solutions, it had explained the dependences of the chain relaxation time on molecular weight and concentration of the polymer, and also the dependence on the light scattering wave-vector [33,37,39-41].

## 4. Discussion on other related dynamics of PEO in blends with PMMA

Mention has been made in the previous sections that the crossover from Rouse dynamics to slowed-down non-Rouse dynamics of PEO chains in blends with PMMA is a special case of a general phenomenon of relaxation and diffusion in many-body interacting systems. At some $t_c$ determined by the interaction potential, the primitive or independent relaxation crosses over to the slowed-down intermolecularly coupled relaxation. It is instructive to give an example of such phenomenon coming from the segmental α-relaxation of the PEO component in blends with PMMA with similar composition as studied in Refs.[6] and [7].

First and foremost, the crossover has been found in the study of segmental α-relaxation of PEO in blends with PMMA. Garcia Sakai et al. [31,32] performed quasielastic neutron scattering measurements of pure PEO, and hPEO in blends with dPMMA with 10, 20 and 30% of hPEO at $Q$=0.89 and 2.51 Å$^{-1}$. At higher temperatures, the intermediate scattering function $F(Q,t)$ of PEO decays principally by the primitive relaxation via exp($-t/\tau_0$) at times before $t_c$≈1 ps and $\tau_0$ is of the order of picoseconds, and the rest after crossing $t_c$ by a slower decaying function. The decay is slower in blends with higher concentration of PMMA, just like in the case of chain relaxation of PEO in blends with PMMA after crossing $t_c$≈1 ns.

Second, we revisit some experimental data on the segmental α-relaxation of the PEO component acquired by deuteron NMR at high frequencies from 31 to 76 MHz by Lutz et al. [3]. It was found that the segmental dynamics of PEO are nearly independent of composition in blends ranging from 0.5 to 30% PEO. The segmental α-relaxation times, $\tau_\alpha$, of the PEO component in these mixtures do not increase much with decreasing PEO concentration, and are hardly influenced by the presence of the considerably slower moving PMMA. For example, $\tau_\alpha$ of 0.5% PEO chains dissolved at near tracer compositions in a PMMA matrix are retarded by less than one order of magnitude when compared with pure PEO samples. The result holds for a wide range of temperatures extending to well below the glass transition of the PMMA matrix, where the segmental relaxation times of PMMA are about 12 orders of magnitude greater than the fast PEO relaxation times. These observations at high frequencies are highly unusual when compared with the component segmental dynamics of other miscible polymer blends measured at a much lower frequencies by dielectric spectroscopy or by another kind of deuteron NMR technique. It was concluded by Lutz et al. that their observations of PEO component dynamics at high frequencies from 31 to 76 MHz cannot be described by current models. Subsequent to the publication of the paper by Lutz et al., the CM [42-45] for polymer blends was applied to explain the data [46]. It was demonstrated that the predictions of the CM on the PEO dynamics in the PEO/PMMA blends from 0.5% to 30% PEO are consistent with the experimental findings that $\tau_\alpha$ of the PEO component is nearly composition-independent over the entire composition range, and is retarded by less than one order of magnitude when compared with $\tau_\alpha$ of pure PEO [3]. The cause of this unusual behavior of $\tau_\alpha$ of the PEO component is due to the high frequencies used in the NMR measurements, resulting in primitive relaxation times $\tau_0$ of the PEO component that are short and not much longer than the crossover time $t_c$≈ 2 ps observed by neutron scattering [31,32] and considered in the CM [46,47]. Usually the segmental α-relaxation of the faster component in other polymer blends was measured by dielectric relaxation or by other NMR techniques, all at times much longer times than 2 ps. In these lower frequency/lower temperature measurements, large increase of $\tau_\alpha$ with enhanced temperature dependence was observed, and the effect increases with increasing presence of the slow component. This normal behavior was found in this case because the primitive relaxation time $\tau_0$ is now much longer than the crossover time $t_c$≈ 2 ps. Quantitative details of the treatment and comparison with experimental data in both cases can be found in Refs.[46,47]. Analogues of these two contrasting situations in segmental α-relaxation can be found for the present problem of PEO chain dynamics in blends with PMMA. Here the primitive relaxation time is $\tau_R(T,Q)$, and the crossover time is 2 ns. We can pick out a case similar to the high frequency NMR measurements of segmental α-relaxation from the data in Fig.4b of Ref.[7]. When $T=350$ K and $\log Q=-0.4$, -0.30, and -0.22 corresponding to $Q=0.4$, 0.5 and 0.6 Å$^{-1}$, it can be seen that $\tau_R(T,Q)$ (open circles in the figure) is less than or equal to $t_c \approx 1$ ns, and the corresponding $\tau(T,Q)$ (closed circles) is only factor ranging from 3 to 5 times longer, and the temperature dependence of the two are slightly different, albeit $\beta(T,Q)=0.34$ at 350 K. On the other hand, for other choices of $T$ and $Q$ such that $\tau_R(T,Q)$ becomes much longer than $t_c \approx 2$ ns, $\tau(T,Q)$ becomes much longer and has a stronger $T$-dependence than $\tau_R(T,Q)$.

## 5. Conclusion

Crossover from primitive or independent relaxation to slowed-down many-body relaxation at some time $t_c$ is a general property of relaxation in interacting systems. It was indeed observed in the dynamics of PEO chain in blends with PMMA by quasielastic neutron scattering and MD simulations. This property alone is sufficient to explain the data without invoking additional assumption, such as done in the Random Rouse Model where each monomer in the Gaussian PEO chain has an individual mobility obeying a broad log-normal distribution and the width is adjustable. In particular, the crossover property by itself can explain the dependence of the relaxation time on momentum transfer $Q$ and temperature $T$, and the trend of the changes on increasing the concentration of PMMA in the blend. The generality of the crossover of dynamics is made more vivid by showing that it also occurs in the segmental α-relaxation of PEO in blends with PMMA of similar compositions as in the present study. Moreover, the anomalous effect observed in high frequency deuteron NMR experiment on the segmental α-relaxation of PEO in blends with PMMA is revisited to show that it is connected with the crossover property. Similar effects are found in the dynamics of PEO chain in blends with PMMA.

**Acknowledgment**

Support of the research came from National Basic Research Program of China (973 Program No. 2010CB731604), and by NSFC (Grant Nos.50731005, 50821001, 10804093, and 51071138). The authors thank Simone Capaccioli for making us aware of the problem discussed in this paper.